\documentclass[preprint,aps]{revtex4}     

\begin{document}


\title{Finite-Temperature Properties of  Ba(Zr,Ti)O$_3$ Relaxors \\ From First Principles}

\author{A. R.  Akbarzadeh$^{1}$, S. Prosandeev$^{2,3}$, Eric J. Walter$^{4}$, A. Al-Barakaty$^{5}$ and L. Bellaiche$^{2}$}

\affiliation{$^{1}$Wiess School of Natural Sciences, Rice University,
6100 Main Street, MS-103, Houston, TX 77005, USA \\
  $^{2}$ Physics Department and Institute for Nanoscience and Engineering,
  University of Arkansas, Fayetteville, Arkansas 72701, USA \\
  $^3$Physics Department and Institute of Physics, South Federal University, RUSSIA\\
                $^{4}$Department of Physics, College of William and Mary, Williamsburg, VA 23187, USA\\
                 $^{5}$Physics Department, Teacher College, Umm Al-Qura University, Makkah, Saudi Arabia
                }

\date{\today}

\begin{abstract}

A first-principles-based technique is developed to investigate properties of
Ba(Zr,Ti)O$_3$ relaxor ferroelectrics as a function of temperature.
The use of this scheme provides answers to important, unresolved and/or controversial questions, such as:
what do the different critical temperatures usually found in relaxors correspond to? Do polar nanoregions really exist in relaxors? If yes, do they only form inside chemically-ordered regions? Is it necessary that antiferroelectricity develops in order for the relaxor behavior to occur? Are random fields and random strains really the mechanisms responsible for relaxor behavior? If not, what are these mechanisms? These {\it ab-initio-based} calculations also leads to a deep microscopic insight into relaxors.

\end{abstract}

\pacs{77.80.Jk, 64.70.Q-, 77.84.-s, 77.80.B-}
\maketitle

%



\marginparwidth 2.7in
\marginparsep 0.5in

Relaxor ferroelectrics are characterized by some striking anomalous properties (see, e.g., Refs.~\cite{Cross,burns83,Smolenskii,Westphal,Tagantsev,Pirc,Egami2005,Bai,Vogel,Fulcher,Brahim,Svitelskiy2005,Tinte,Sobolev2005,Takesue1999,blinc,Vugmeister,Viehland,Colla,Rappe,Jirka} and references therein).  For instance, they adopt
a peak in their $ac$ dielectric response-versus-temperature function while they remain macroscopically paraelectric and cubic down to the lowest temperatures  \cite{Cross}. Furthermore, this dielectric response deviates from the ``traditional'' Curie-Weiss law \cite{CW} for temperatures lower than the so-called Burns temperature \cite{burns83}. Other examples of anomalous properties include the plateau observed in their static, $dc$ dielectric response at low temperature \cite{Kutnjak,Levstik}, and the unusual temperature behavior  \cite{blinc} of the Edwards-Anderson parameter  \cite{Edwards}. Determining the origin of these intriguing effects has been a  challenge to scientists since the discovery of ferroelectric relaxors.

The goal of this Letter is to report {\it ab-initio}-based calculations that not only reproduce all the aforementioned intriguing features but also offer a deep microscopic insight into relaxors.

Practically, we decided to focus on a specific relaxor, namely disordered  Ba(Zr$_{0.5}$Ti$_{0.5}$)O$_3$ (BZT) solid solutions (BZT is also fascinating because its parent compounds are rather different: BaZrO$_3$ is paraelectric while BaTiO$_3$ is a typical  ferroelectric).
Here, we develop and use a first-principles-based effective Hamiltonian approach for which
a detailed description is given in the Supplemental Material.
 The total energy of this effective Hamiltonian is used in Monte-Carlo (MC) simulations to compute finite-temperature properties of BZT alloys. We use $12\times 12\times 12$ (8640 atoms) or $16\times 16\times 16$ (20480 atoms) supercells in which the $\sigma_j$  variables  are randomly placed and kept fixed during the MC simulations, in order to mimic disordered BZT solid solutions. These two supercells provide similar results, which attest the convergency of the simulations.
The temperature $T$ is decreased in small steps from high temperature, and  up to 10$^{6}$ MC sweeps are used to get converged statistical properties.

Here, the ${\bf u_{\it i}}$ local soft-mode vectors in each 5-atom cell $i$ (${\bf u_{\it i}}$ is directly proportional to the local electric dipole moment centered in cell $i$) and  the homogeneous strain tensor $ \eta_{\it H}$ arising from the MC simulations indicate that  Ba(Zr$_{0.5}$Ti$_{0.5}$)O$_3$ bulk remains macroscopically {\it cubic and non-polar} for {\it any} temperature down to the lowest one investigated here (which is 5\,K), as consistent with measurements \cite{Bhalla}.
We also computed the dielectric susceptibility, at different temperatures  by progressively cooling down the system, from our MC simulations via two different approaches: (i) a ``direct'' method for which the resulting dielectric susceptibility is denoted as
$\chi^{direct}$ and is calculated as the change in polarization with respect to an applied electric field (with this field practically being oriented along the [111] pseudo-cubic direction and having a magnitude of 10$^7$ V/m); and (ii) the ``correlation-function'' approaches of Refs.~\cite{Alberto1,Karin} for which the resulting dielectric susceptibility is referred to as $\chi^{CF}$ and is provided by the fluctuation-dissipation theorem via $\chi^{CF}_{\alpha \beta}= \frac{(N~Z^{*})^2}{V\epsilon_o k_BT} \left [ < u_{\alpha} u_{\beta} > - < u_{\alpha}> < u_{\beta} >  \right ]$,
where $< u_{\alpha} u_{\beta}>$ denotes the statistical average of the product between the  $\alpha$ and $\beta$ components of  the supercell average of the  local mode vectors, and where $< u_{\alpha}>$  (respectively, $< u_{\beta}>$) is the statistical average of the $\alpha$- (respectively, $\beta$-) component of the supercell average of the  local mode vectors.
$N$ is the number of sites in the supercell while $V$ is its volume.  $k_B$ is the Boltzmann's constant and $\epsilon_o$ is the permittivity of the vacuum. Strikingly, while previous works (see, e.g., Ref. \cite{Karin}) found that these two different methods provide nearly identical dielectric susceptibilities in typical ferroelectrics, Fig. 1(a) reveals that it is not the case for disordered BZT:  $\chi^{CF}$ exhibits a peak around  T$_{f}$ $\simeq$ 130K, while $\chi^{direct}$ increases when decreasing the temperature down to T$_{f}$  and then saturates to a plateau for lower temperature. Both the temperature behavior of $\chi^{CF}$ and the temperature at which $\chi^{CF}$ is maximum
are fully consistent with the dielectric experiments of Ref. \cite{Bhalla} in Ba(Zr$_{x}$Ti$_{1-x}$)O$_3$ {\it relaxors}  under $ac$ electric fields having frequencies ranging between 100Hz and 100kHz. Moreover, the depicted behavior of $\chi^{direct}$ is exactly the one expected for the perfectly {\it static} dielectric response of relaxors \cite{Kutnjak,Levstik}, which allows us to identify T$_{f}$ as the so-called freezing temperature \cite{Egami2005,Bai,Vogel,Fulcher} (a freezing temperature ranging between 100 and 140K has been reported for BZT systems \cite{Farhi1999}, in good agreement with our value of $\simeq$ 130K).
Our $\chi^{direct}$ thus provides the static (dc) dielectric response while our simulated $\chi^{CF}$ corresponds to observed low-frequency dielectric responses of BZT relaxors -- which is reminiscent of the fact  that the susceptibility given by the  fluctuation-dissipation theorem is nearly the $ac$-susceptibility in the Edwards-Anderson model of spin glasses \cite{Mydosh}.

It is also important to recall that, while  $\chi^{CF}$ possesses a peak at T$_f$, our MC simulations indicate that Ba(Zr$_{0.5}$Ti$_{0.5}$)O$_3$ bulk remains macroscopically {\it cubic and non-polar} for {\it any} temperature -- as consistent with what is expected for relaxors \cite{Cross}.
 Moreover, the temperature behaviors of $\chi^{CF}$ and  $\chi^{direct}$ allow the introduction of four different regions, namely (1) Region I that concerns temperatures, $T$,  above T$_{b}$ $\simeq$ 450\,K  and for which $\chi^{CF}$ and  $\chi^{direct}$ can be nicely fitted by the Curie-Weiss formula \cite{CW}, i.e. they are both directly proportional to $1/ |T-T_0|$ (where $T_0$ is practically found here to be very close to $-$120K);  (2) Region II that extends between T$^{*}$ $\simeq$ 240\,K  and T$_{b}$ for which $\chi^{CF}$ increases as the temperature decreases but does not follow anymore the Curie-Weiss law, unlike $\chi^{direct}$; (3) Region III that is located in-between T$_f$ and  T$^{*}$ for which neither  $\chi^{CF}$ nor  $\chi^{direct}$ obey the Curie-Weiss law; and Region IV that occurs for temperature lower than T$_f$, and for which $\chi^{CF}$ decreases as $T$ is reduced while $\chi^{direct}$ is nearly constant there. T$_{b}$ can be assigned to be the Burns temperature \cite{burns83} while T$^{*}$ can be thought as being the novel critical temperature recently found in relaxors \cite{Brahim,Svitelskiy2005}. The facts that $\chi^{CF}$ follows the Curie-Weiss law only for temperatures above the Burns temperature and that this Burns temperature is of the order of 450K have both been observed in Ba(Zr$_{0.5}$Ti$_{0.5}$)O$_3$ \cite{Bhalla}. Similarly, a {\it negative}  $T_0$ Curie temperature   has also been experimentally extracted in BZT samples \cite{Bhalla}.

Figure 1(b) reports the temperature evolution of the so-called Edwards-Anderson (E-A) parameter \cite{Edwards},  $q_{EA}$, that is calculated as
 $q_{EA}=\left\langle  {\left\langle {Z^* {\bf u_{\it i}}} \right\rangle _t }
^2 \right\rangle _{i}$, where the inner averaging is made over the $t$ Monte-Carlo sweeps while the outer averaging is
made over the $i$ lattice sites.
The behavior of the simulated $q_{EA}$  of BZT bulk versus temperature  bears some striking resemblance with those  predicted by the spherical random-bond-random-field model and measured from nuclear magnetic resonance for PbMg$_{1/3}$Nb$_{2/3}$O$_3$ relaxor \cite{blinc}. For instance, (1) it is small and nearly linearly increases when decreasing the temperature at large temperature (in Region I); (2) it is large and also linearly increases when decreasing the temperature for small temperature (in Region IV); and (3) the $q_{EA}$-{\it versus}-$T$ function is curved upward in-between (in Regions II and III).
Figure 1(b) also reveals that the temperature behavior and values of the overall  Edward-Anderson parameter  (for any temperature) almost entirely originate from the electric dipoles centered on Ti ions. Consequently, the contribution of the dipoles belonging to BaZrO$_3$ unit cells on the total Edwards-Anderson parameter nearly vanishes. Other dramatic differences between local properties associated with Zr {\it versus} Ti atoms are reported in Fig. 1(c), which shows that not only the average magnitude of the local dipoles centered on Zr ions is much smaller than those centered on Ti ions, but its temperature behavior is also strikingly different: the dipoles belonging to BaZrO$_3$ unit cells continuously shrink in average as the temperature is reduced, while the dipoles located inside BaTiO$_3$ cells suddenly become enlarged when decreasing the temperature below T$^{*}$. Electric diffraction measurements \cite{Liu2007} and a model emphasizing the importance of the BaTiO$_3$ soft mode on the relaxor behavior of BZT \cite{Simon2004} are also consistent with our prediction that the Ti sites carry much larger dipoles  than Zr sites. Moreover, the results from Fig. 1(c) imply that, at the lowest temperatures, the Ti atoms displace in average by about 0.16 \AA, while the Zr atoms move by 0.03 \AA~ from their cubic, equilibrium positions. Such numbers are in remarkable agreement with the values of 0.17 and 0.03 \AA, respectively, obtained by the first-principles calculations of Ref. \cite{Pasturel} for a BZT supercell containing 135 atoms \cite{footnotelength}.

Let us now focus on Figures 2, that display dipolar snapshots within a given (y,z) plane at different temperatures, in order to gain a microscopic understanding of relaxors.
Figure 2a reveals that Region I consists of randomly oriented dipoles that are centered on Ti ions and that are surrounded by much smaller dipoles located inside BaZrO$_3$ cells.
As indicated by Fig. 2b, some of these Ti sites act as nuclei to the formation of small clusters inside which the dipoles begin to be parallel to each other in Region II. We numerically found that the polarization of these small clusters in Region II does not automatically lie along a $<111>$ direction. For instance, the average direction of the local modes inside the bottom cluster of Fig. 2b is along an orthorhombic-like direction, namely it is close to $[01\bar{1}]$. Interestingly, some of these clusters do not even possess a polarization being parallel to a high-symmetry direction in Region II, such as the top cluster of  Fig. 2b for which the vector resulting from the average of the local modes is equal to (-0.012,-0.052,-0.021) in the (x,y,z) basis -- that is a triclinic direction. It is interesting to realize that thermal strain measurements \cite{Bhalla} strongly suggest that polar nanoclusters can exist in BZT up to $\simeq$ 440K, which is consistent with our finding of small polar clusters in Region II (that extends up to  T$_{b}$ $\simeq$ 450\,K).

As the system enters Region III, two novel features occur as it can been inferred from Fig. 2c. First of all, more (small) polar clusters form as the temperature is decreased, which makes the average magnitude of the Ti dipoles increasing (see Fig.  1c). Secondly, some of these clusters now possesses a polarization being close to a  $<111>$ direction, such as the left and right clusters displayed in Fig. 2c for which the average local mode is equal to (0.043,-0.048,0.043) and (0.034,0.037,0.045), respectively. Note that, while the clusters are always formed at Ti sites, they do not necessarily stay at the same sites for different temperatures, or even for different MC sweeps at the same temperature, in Regions II and III. In that sense, they can be thought of being of dynamical nature rather than being static.

Below T$_f$, some of these clusters have considerably grown in size, like the one located at the bottom right corner in Figs. 2 d-f.   Novel clusters can still form when decreasing the temperature in Region IV, such as the one near the bottom left corner of Fig. 2f at 10K. On the other hand, other clusters are frozen in the sense that they are always located at the same region of space and have a polarization that lies along the same direction, independently of the temperature and MC sweep in Region IV (see the central and bottom right clusters in Figs. 2d-f). While the different clusters possess different numbers of Ti sites and have different overall shapes, they share a common feature in Region IV: they all have a polarization being close to one of the eight equivalent   $<111>$ directions, as  consistent with the experimental finding that Raman spectra indicate a rhombohedral structure for the polar regions at liquid nitrogen temperature in BZT relaxors \cite{Dixit2006}. As the temperature is reduced in Region IV, the matrix  possesses Zr-centered dipoles that are significantly shrinking in magnitude. This matrix in Regions  II,  III  and IV also possesses individual Ti dipoles that are oriented along many different directions, as in Region I.

To have further insight into the relaxor behavior,  let us denote as {\bf k$_{max}$} the vector
of the first Brillouin zone possessing the largest magnitude of the Fourier transform of the local dipoles configuration \cite{aaronjorge}. {\bf k$_{max}$} is numerically found to be slightly dependent on the choice of the used supercell, but is always a non-highly symmetric vector that is close to neither the center nor the boundary of the cubic first Brillouin zone. For instance, in case of a $12\times 12\times 12$ supercell,  ${\bf k_{max}}$= $ \frac{2\pi}{6 a_{lat}} (-{\bf y} + {\bf z})$, where $a_{lat}$ is the lattice constant of the 5-atom primitive cell and where {\bf y} and {\bf z} are unit vectors along the y- and z-axis, respectively.
Figure 1(d) reports the temperature evolution of the square of the Fourier transform of the local dipoles configuration at {\bf k$_{max}$}. One can clearly see that, in Regions I and  II, this quantity is nearly zero. On the other hand, it increases when the temperature decreases below T$^*$  while still remaining fairly small (around 1.5\% of the total spectra gathering the Fourier transforms at all possible k-points, at 5K). We interpret such latter results as indicative that the different nanopolar regions slightly interact  in Regions III and IV  in an antiferroelectric-like (or incommensurate \cite{Uesu}- or dipolar-wave-like) fashion. Interestingly, antiferroelectricity has been previously reported in some relaxor systems \cite{Sobolev2005,Takesue1999}.

Let us now compute the correlation between Ti dipoles (we decided to focus on Ti-Ti dipolar correlations because  Figs. 2 revealed that the polar clusters only contain Ti sites and because  Fig. 1b shows that the overall Edwards-Anderson parameter mostly only originates from Ti dipoles).  This correlation is  practically defined by $\theta ({\bf r}) =\frac{1}{N_{Ti}} \sum_{i}  \frac{{\bf u_i} \cdot {\bf u_{i+r}}}{\left|{\bf
u_i}\right|  \left|{\bf u_{i+r}}\right|}$, where the index $i$ runs over all the $N_{Ti}$ Ti-sites of the system and where ${\bf u_i}$ and
$ {\bf u_{i+r}}$ are the local modes in cell $i$ and in the cell centered on the Ti atom (if any) distant from
${\bf r}$ from the cell $i$, respectively \cite{AliKTO}. A value of  1 (respectively, -1) for $\theta ({\bf r})$ for a given ${\bf r}$ would indicate that Ti dipoles and their neighboring Ti dipoles distant from ${\bf r}$ are aligned along  the same (respectively, opposite) direction. Figure 1(e) shows the value of $\theta ({\bf r})$  for various representative ${\bf r}$ vectors, as a function of temperature. One can see that, in Region I and in average, the Ti dipoles are only (and slightly) correlated with the Ti dipoles centered at their first nearest neighboring cells. Such correlation further increases in strength when the polar nanoclusters form and become bigger in size and in polarization, as the temperature is reduced in Regions II, III and IV. Second and third-nearest neighbors also begin to be more correlated in average as the temperature decreases in Regions III and IV. Interestingly,  a significant {\it anticorrelation}  (see the negative sign of the correlation) between Ti dipoles being distant by 3 lattice constant along the z- (or x- or y-) axis also strongly develops in Regions III and IV, which reinforces the previous finding that antiferroelectric-like interactions exist within the BZT relaxor system. Note that the Supplemental Material also provides
and discusses the $\theta ({\bf r})$ function for all the ${\bf r}$-vectors lying in the (y,z) plane at 10K.

A particularly important feature of our scheme  is that we can switch on and off some interactions in order to determine their effect on physical properties. We numerically found that turning off random fields and random strains does {\it not} significantly affect the results shown in Figs 1-2, which contrasts with a  common belief on the microscopic origins of relaxors \cite{blinc,Vugmeister} while being more consistent with models proposed for the homovalent (K,Li)TaO$_3$ relaxor \cite{ToulouseKLT,VugmeisterKLT}. On the other hand, our computations reveal that it is the difference in polarizability between Ti and Zr ions that leads to the relaxor behavior in BZT. As a matter of fact,  annihilating such differences in the simulations leads to (1) $\chi^{direct}$ and $\chi^{CF}$ being equal to each other and continuously decreasing as the temperature decreases down to 0K (with the system remaining cubic and non-polar); (2) the Edwards-Anderson parameter being around ten times smaller than the one depicted in Fig. 1b at low temperature, and (3) the polar nanoclusters disappearing. It should also be emphasized that our simulations results depicted in Figs 1-2 imply that
relaxor behavior can occur in BZT even if  no large chemically-ordered region exists in that system  (since our computations were performed on disordered solid solutions). Such finding seems to contrast with models recently proposed to explain the relaxor behavior of heterovalent Pb(Sc,Nb)O$_3$ and Pb(Mg,Nb)O$_3$ alloys \cite{Tinte}, while agreeing with a study   \cite{Liu2007} downplaying the role of chemical short-range ordering on the formation of polar nanoregions in BZT.
In fact, our simulations indicate that the relaxor behavior already occurs in {\it disordered} BZT solid solutions because some regions of space can be more Ti-rich  than others because of the random process of assigning sites in a disordered solid solution. Such feature bears resemblance with the Anderson localization phenomenon for which electronic wave-functions become  localized in a region of space (of an overall disordered $(A',A'')$ solid solution) being much richer in $A'$ than in $A''$ \cite{Linwang}. Finally, we also increased the antiferroelectricity-like interactions (by playing with the so-called j$_5$ short-range coefficient \cite{ZhongDavid}). We found that such increase leads to a shift towards higher temperature of the peak of $\chi^{CF}$,  in addition to enhance at low temperature
(i) the  Edwards-Anderson parameter, (ii) the average magnitude of the local modes centered on Ti ions, (iii) the  square of the Fourier transform of the local dipoles configuration at {\bf k$_{max}$} and (iv) the strength of the anticorrelation between Ti dipoles being distant by 3 lattice constants along the z- (or x- or y-) axis. Such findings emphasize the importance of the antiferroelectricity-like interactions between Ti-rich nanopolar clusters for the relaxor behavior.

We therefore hope that our study helps in better understanding the fascinating relaxor ferroelectrics. In order to further enhance such understanding, future studies could examine the influence of static and dynamic (GHz-THz) electric fields \cite{Rappe,Jirka} on the behaviors of BZT materials, and  determine if the results found here also hold for heterovalent relaxors (such as Pb(Sc,Nb)O$_3$ and Pb(Mg,Nb)O$_3$).

This work is financially supported by the ONR Grants N00014-11-1-0384 and N00014-08-1-0915. S.P. and L.B. also acknowledge the  NSF DMR-1066158 and DMR-0701558, Department of Energy, Office of Basic Energy Sciences, under contract ER-46612, and ARO Grant W911NF-12-1-0085 for discussions with scientists sponsored by these grants. Some computations  were also made possible thanks to the ONR grant N00014-07-1-0825 (DURIP), MRI grant 0722625 from NSF, and a Challenge grant from the Department of Defense. S.P. appreciates Grant 12-08-00887-a of Russian Fund for Basic Research.
The authors thank Drs A. Bhalla, Igor Kornev and Sergey Lisenkov for  discussions.



\newpage

FIGURE CAPTIONS

\vspace{5mm}

Figure 1: Temperature dependency of some properties in disordered Ba(Zr$_{0.5}$Ti$_{0.5}$)O$_3$ solid solutions. Panel (a) shows the average between the three diagonal elements of the dielectric susceptibility, as computed from a direct approach ($\chi^{direct}$, triangles) and from the fluctuation-dissipation theorem ($\chi^{CF}$, dots). Panel (b) displays the total Edwards-Anderson parameter, as well as its contributions from cells centered on Ti and Zr ions. Panel (c) reveals the
magnitude of the local modes centered on Ti and Zr ions. Panel (d) represents the
square of the Fourier transform of the local modes' configurations at {\bf k$_{max}$}. Panel (e) provides
the $\theta ({\bf r})$ correlation between Ti dipoles for ${\bf r}$= $a_{lat} {\bf z}$ (first nearest neighbor),
$a_{lat} ({\bf y}+{\bf z})$ (second nearest neighbor), $a_{lat} ({\bf x}+{\bf y}+{\bf z}$) (third nearest neighbor), $2a_{lat} {\bf z}$ and $3a_{lat} {\bf z}$. The solid line in Panel (a) represents the dielectric susceptibility arising from the fit of  $\chi^{CF}$ (between 500 and 800K) by the Curie-Weiss law  \cite{CW}.

\vspace{5mm}

Figure 2: Snapshots of the dipolar configurations in a given (y,z) plane for different temperatures.
Panels (a), (b), (c), (d), (e) and (f) correspond to temperature of 550K (Region I), 250K (Region II),  150K (Region III), 100K (Region IV), 50K (Region IV) and 10K (Region IV), respectively.
Blue colors and red colors indicate that the corresponding local modes are centered on Ti and Zr ions, respectively.

\newpage

\begin{thebibliography}{999}
\bibitem{footnotelength} Note that, in order to determine the actual atomic displacements of Ti and Zr atoms from our MC simulations, one has to multiply the results of Fig. 1(c) by the product of the lattice constant with the eigenvector of the force-constant matrix associated with Zr or Ti atoms. This product is of the order of 2.55 \AA.
\end{thebibliography}
\end{document}



\title{Supplemental Material of \\ Finite-Temperature Properties of  Ba(Zr,Ti)O$_3$ Relaxors \\ From First Principles }

\author{A. R.  Akbarzadeh$^{1}$, S. Prosandeev$^{2,3}$, Eric J. Walter$^{4}$, A. Al-Barakaty$^{5}$ and L. Bellaiche$^{2}$}
\affiliation{$^{1}$Wiess School of Natural Sciences, Rice University,
6100 Main Street, MS-103, Houston, TX 77005, USA \\
  $^{2}$ Physics Department and Institute for Nanoscience and Engineering,
  University of Arkansas, Fayetteville, Arkansas 72701, USA \\
  $^3$Physics Department and Institute of Physics, Southern Federal University, Russia \\
                $^{4}$Department of Physics, College of William and Mary, Williamsburg, VA 23187, USA\\
                 $^{5}$Physics Department, Teacher College, Umm Al-Qura University, Makkah, Saudi Arabia}

\maketitle

%





\section{ First-principles-based effective Hamiltonian approach}

Here, we develop a first-principles-based effective Hamiltonian approach for which the
the total energy, $E_{tot}$, of Ba(Zr,Ti)O$_3$ (BZT) solid solutions is written as a sum of two main terms:
\begin{eqnarray}
   E_{tot} (\{ { \bf u_{\it i}} \}, \{ { \bf v_{\it i}} \}, \eta_{\it H},
       \{ \sigma_{\it j} \}) =
   E_{\rm ave} (\{ { \bf u_{\it i}} \}, \{ { \bf v_{\it i}} \}, \eta_{\it H})
   +~E_{\rm loc} (\{ { \bf u_{\it i}} \},\{ { \bf v_{\it i}} \},
   \{ \sigma_{\it j} \}) \;\;,
\end{eqnarray}
where ${\bf u_{\it i}}$ is the local (Zr or Ti-centered) soft-mode in unit cell
$i$. Its product with its associated Born effective charge provides the local electric dipole moment centered in cell $i$. $\{ { \bf v_{\it i}} \}$ are Ba-centered dimensionless local displacements that are related to the inhomogeneous strain variables inside each cell \cite{ZhongDavid}, while $\eta_{\it H}$ is the homogeneous strain tensor \cite{ZhongDavid}. Finally, $\{ \sigma_{{\it j}}\}$ represents the atomic configuration of the BZT solid solution.  Practically,  $\sigma_{\it j}$=+1 or $-1$ indicates the
presence of a Zr or Ti atom located at the  lattice site $j$, respectively.
The first energetic term of Eq. (1), $E_{\rm ave}$, describes the interactions in a Ba$<B>$O$_3$ virtual crystal system, where $<B>$ is a virtual atom involving a kind of potential average between Zr and Ti atoms~\cite{LaurentDavid3}. The second energetic term, $E_{\rm loc}$, describes how the actual distribution of Zr and Ti cations affects the energetics involving the local soft-modes ${\bf u}_i$  and the local strain variables, and therefore depends on the $\{\sigma_j \}$ distribution.
Practically, the analytical expression of $E_{\rm ave}$ is provided in Ref.[\onlinecite{ZhongDavid}], and contains five energetic terms:  a local mode self-energy, long-range and short-range interactions between local modes, an elastic energy and interactions between local modes and strains.
$E_{\rm loc}$ is proposed to be given by:
\begin{eqnarray}
  E_{\rm loc} && (\{ { \bf u_{\it i}} \},\{ { \bf v_{\it i}} \},
  \{ \sigma_{\it j} \}) = \nonumber \\
  && \sum_{i} [ \Delta \kappa (\sigma_{\it i}) ~ u_{\it i}^{2} ~+\Delta \alpha (\sigma_{\it i}) ~ u_{\it i}^{4} ~+
  ~ \Delta \gamma (\sigma_{\it i}) ~( u_{\it ix}^{2}u_{\it iy}^{2} +
  u_{\it iy}^{2}u_{\it iz}^{2} + u_{\it iz}^{2} u_{\it ix}^{2})]\nonumber \\
  && +~ \sum_{ij}
  [Q_{\it |j-i|}~\sigma_{\it j}~ { \bf e_{\it ji}} \cdot { \bf u_{\it i}}~+~
  R_{\it |j-i|}~\sigma_{\it j}~ { \bf f_{\it ji}} \cdot { \bf v_{\it i} }]
  \;\;,
\end{eqnarray}
where the sum over $i$ runs over all the unit cells, while the sum
over $j$ runs up to the third-nearest neighbors of cell $i$.
$u_{\it i \beta}$, with $\beta=$x, y or z, denote the Cartesian coordinates of
the local-mode $\bf{u_{\it i}}$.  $\bf{e_{\it ji}}$ is a unit vector
joining the site $j$ to the center of the soft mode $\bf{u_{\it
i}}$, and $\bf{f_{\it ji}}$ is a unit vector joining the site $j$
to the origin of $\bf{v_{\it i}}$.  $\Delta \kappa (\sigma_{\it i})$
characterizes the on-site {\it harmonic} contribution of alloying, while
$\Delta \alpha (\sigma_{\it i})$ and $\Delta \gamma (\sigma_{\it i})$ represent  the effect of
alloying on {\it anharmonicity}.
$Q_{{\it |j-i|}}$ and $R_{{\it |j-i|}}$ quantify the
intersite interactions between the atomic variable $\sigma_{\it
j}$ on the site $j$ and the local mode $u_{\it i}$ and the
strain-related $v_{\it i}$ at the site $i$, respectively.
The $Q_{{\it |j-i|}}$ and $R_{{\it |j-i|}}$ parameters can be considered as characterizing the strengths of the random electric fields and random strain fields in disordered solid solutions, respectively \cite{PSNrelaxor}.
Note that Eq. (2) is similar to the expression previously proposed for Pb(Zr,Ti)O$_3$
solid solutions \cite{PRLPZT}, at the important exception that the $\Delta \kappa$ parameters of Zr and Ti are presently different here. Such difference can be considered as characterizing the difference in polarizability between BaZrO$_3$ and BaTiO$_3$ \cite{Sergeyembedded}, and, is found to be crucial
here to reproduce and to understand relaxor behavior in BZT.

All the parameters entering the analytical expressions of Eqs. (1) and (2) are obtained from
first-principle calculations \cite{LDA,USPP,LaurentDavid3} on small cells. The total energy of Eq.(1) is then used in Monte-Carlo (MC) simulations to compute finite-temperature
properties of BZT alloys. We use $12\times 12\times 12$ (8640 atoms) or $16\times 16\times 16$ (20480 atoms) supercells in which the $\sigma_j$  variables  are randomly placed and kept fixed during the MC simulations, in order to mimic disordered BZT solid solutions. Note that these two supercells provide similar results, which attest the convergency of the simulations.
The temperature $T$ is decreased in small steps from high temperature, and  up to 10$^{6}$ MC sweeps are used to get converged statistical properties.

Note that first-principles-based effective Hamiltonians \cite{PRLPZT,17-kornev-pzt,Walizer2006} have been previously developed and used to study several complex phenomena in different ferroelectric alloys, and have been shown to yield various subtle properties that are in excellent agreement with measurements. Examples include the occurrence of low-symmetry phases and large piezoelectric responses in the so-called morphotropic phase boundary region of Pb(Zr,Ti)O$_3$ systems \cite{PRLPZT,17-kornev-pzt}, as well as phase diagrams, temperature-gradient-induced polarization, and the existence of two modes contributing to the GHz-THz dielectric response of (Ba,Sr)TiO$_3$ materials
\cite{Walizer2006,Quingteng2010,Inna2008,Hlinka2008}.

\begin{figure}[htbp]
\begin{center}
\resizebox{0.88\textwidth}{!}
{\includegraphics{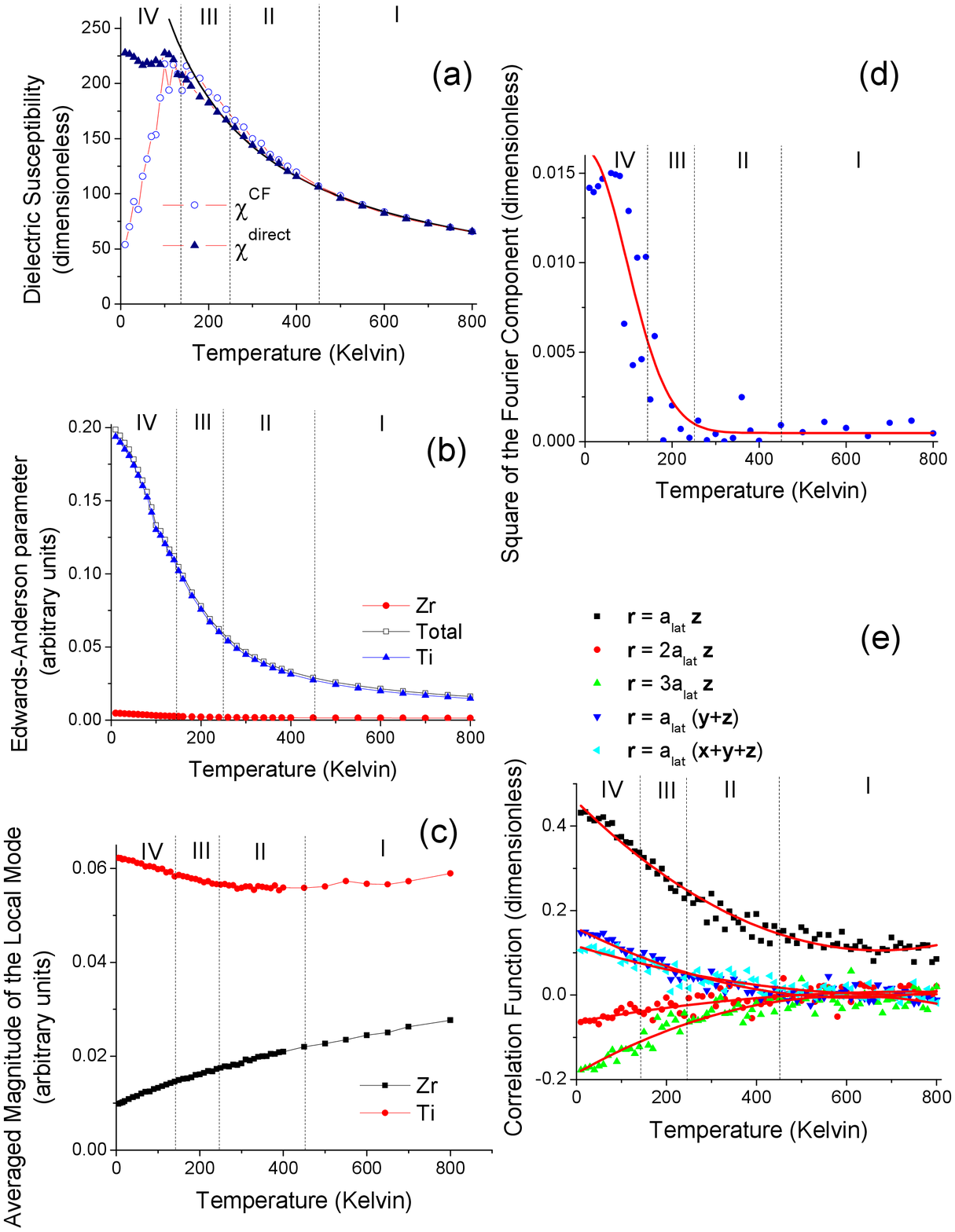}}
\caption{$\theta ({\bf r})$ correlation between Ti dipoles for the ${\bf r}$-vectors lying in the (y,z) plane at 10K (Region IV).}
\end{center}
\label{fig1}
\end{figure}

\section{ Correlation function}

To gain more insight into this Ti-Ti dipolar correlation, Figure 1 of this Supplemental Material displays the $\theta ({\bf r})$ correlation function (defined in the manuscript) for all the ${\bf r}$-vectors lying in the (y,z) plane at 10K. It is worthwhile to realize that  increasing the distance along any  $<001>$ axis leads to the following behavior: $\theta ({\bf r})$  is first positive and strong  (for  ${\bf r}=a_{lat} {\bf z}$), then is slightly negative (for ${\bf r}=2 a_{lat} {\bf z}$), then is significantly negative (for ${\bf r}=3a_{lat} {\bf z}$) and then is much smaller for larger  ${\bf r}$. Figure 1 of this Supplemental Material   also reveals that the Ti-Ti correlation is anisotropic, since, e.g., $<011>$ directions have almost vanishing values of $\theta ({\bf r})$ for any {\bf r} (except for ${\bf r}=a_{lat} ({\bf y}+{\bf z})$) -- unlike
the $<001>$ directions. Our results are therefore consistent with the  dipolar anticorrelation along $<001>$ directions and the anisotropy of the overall correlation  extracted from measurements in BZT  \cite{Liu2007}.